\def\year{2020}\relax
\documentclass[letterpaper]{article} % DO NOT CHANGE THIS
\usepackage{aaai20}  % DO NOT CHANGE THIS
\usepackage{times}  % DO NOT CHANGE THIS
\usepackage{helvet} % DO NOT CHANGE THIS
\usepackage{courier}  % DO NOT CHANGE THIS
\usepackage[hyphens]{url}  % DO NOT CHANGE THIS
\usepackage{graphicx} % DO NOT CHANGE THIS
\usepackage{graphicx}  % DO NOT CHANGE THIS
\title{Does Speech enhancement of publicly available data help build robust Speech Recognition  Systems?}
\author{
Bhavya Ghai,\textsuperscript{\rm 1} 
Buvana Ramanan,\textsuperscript{\rm 2} 
Klaus Mueller\textsuperscript{\rm 1}\\
  \textsuperscript{\rm 1}Department of Computer Science , Stony Brook University, USA\\
  \textsuperscript{\rm 2}Nokia Bell Labs, Murray Hill, USA\\
  \{bghai,mueller\}@cs.stonybrook.edu, buvana.ramanan@nokia-bell-labs.com\\
}
 
\begin{document}

\maketitle

\begin{abstract}
Automatic speech recognition(ASR) systems play a key role in many commercial products including voice assistants. Typically, they require large amounts of clean speech data for training which gives an undue advantage to large organizations which have tons of private data. In this paper, we have first curated a fairly big dataset using publicly available data sources. Thereafter, we tried to investigate if we can use publicly available noisy data to train robust ASR systems. We have used speech enhancement to clean the noisy data first and then used it together with it's cleaned version to train ASR systems. We have found that using speech enhancement gives 9.5\% better word error rate than training on just noisy data and 9\% better than training on just clean data. It's performance is also comparable to the ideal case scenario when trained on noisy and it's clean version.     
\end{abstract}

\section{Introduction}
Automatic speech recognition(ASR) can be understood as a process to convert audio signal to text. ASR systems are a critical part of all voice assistants like siri, cortana, etc. Technology giants like Google, Amazon, etc. leverage tons of private data to build state-of-the-art ASR systems. This makes it really difficult for other players to reproduce similar performance. In this paper, we are trying to investigate if we can use publicly available data to train ASR systems which can compete with the state of the art. If true, it will empower startups, academics, etc. to build competent ASR systems. Publicly available speech data like youtube may be contaminated with ambient noise and background music which makes it difficult to be used for training ASR systems. Hence, we propose speech enhancement techniques to clean the noisy data first and then use the original and it's enhanced(cleaned) version to train ASR systems.       

Speech enhancement(SE) is a well studied problem which aims to enhance audio quality by getting rid of contamination's such as white noise, background music, etc. 
%Existing literature for speech enhancement can be classified as denoising and source separation. In the case of denoising, input audio is considered as a mixture of speech and noise and the objective is to retrieve the speech signal. In case of source separation, input audio might be a blend of many signals and the objective is to retrieve each of the signals accurately. In this study, we will focus on denoising. 
Different GAN based models like SEGAN, FSEGAN, etc. have been shown to perform well for speech enhancement. In this work, we have used SEGAN \cite{segan} which operates at waveform level to remove noise from given noisy speech signal. 
%The generator network in the GAN comprises of strided convolutional layer based auto-encoder whihc performs the enhancement. The discriminator provides feedback to the generator in the form of loss which helps generator tilt towards realistic distribution.
SEGAN uses CNNs instead of RNNs for it's encoder and decoder modules which makes it faster. It operates end to end with raw audio signal so it's free of any assumptions made for feature extraction. Lastly, authors have also shared its code which makes it more reproducible. Hence, we chose SEGAN over other speech enhancement techniques. 
%Speech enhancement and speech enhancement specifically for ASR systems are not exactly the same problems. It's possible that a speech enhancement algorithm which performs well on some SE evaluation metrics mightn't perform as good for ASR systems. So, its   

\begin{figure}[ht]
\centering
\includegraphics[scale=0.40]{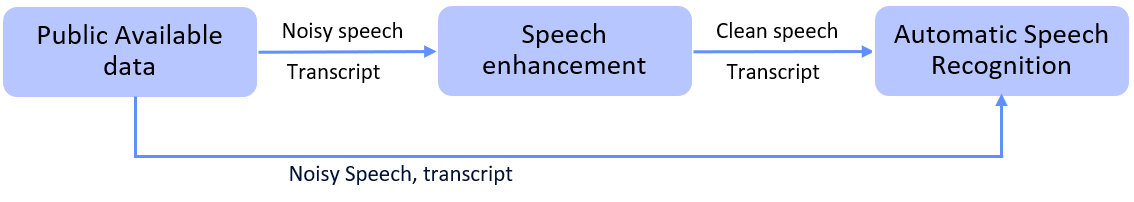}
\setlength{\abovecaptionskip}{-3pt}
\caption{Our Approach: We use publicly available noisy data and it's cleaned version to train ASR model. }
\label{fig:approach}
\end{figure}

There are different processing stages to go about building noise robust ASR systems. Deep learning approaches to build robust ASR systems can be classified into 3 groups i.e. front-end, back-end \& joint front- and back-end techniques\cite{zhang2018deep}. In the front-end setting, speech enhancement and recognition system are independent from each other. Noisy speech is first enhanced during pre-processing and then recognizer is trained on the enhanced speech. In the back-end setting, noisy speech is fed in as it is and the recognizer is optimized such that it outputs the correct corresponding text. Lastly, In joint front- and back-end setting, speech enhancement and recognizer are considered as a single block and trained end-to-end. In this work, we have focused on the front-end and back-end approach as shown in Fig. \ref{fig:approach}.   

One of the most popular approach for back-end setting is multi-condition training. Multi-condition training is a technique which helps make more robust recognition system by training on multiple acoustic variants of the training dataset. 
%One of its earliest applications was for word recognition where training data consisted of normally spoken word along with variants representing the same word but in a fast, clear, loud, etc. variants \cite{multiStyle}. 
In our case, we propose to use publicly available noisy speech along with it's cleaned variant(via SE) for building ASR systems.
% explain a bit about each approach

\section{Dataset}
Existing datasets for speech enhancement are pretty limited in size. ASR systems trained on such datasets  mightn't generalize to different real world conditions. Hence, we decided to curate our own dataset. For clean speech, we used LibriSpeech dataset \cite{panayotov2015librispeech} which is derived from public domain audiobooks. This dataset is fairly big($\sim$460hrs) and its corresponding transcript is also available which makes it suitable for ASR training. Next, we used diverse set of background music and ambient noises to simulate different real world conditions. For ambient noise, we used popular datasets like Urbansound, ESC50 along with youtube. From youtube, we cherry picked videos reflecting background noises in train, traffic, restaurant, rain, etc. For background music, we used youtube to extract movie theme songs and instrumental music belonging to different genres like Latin, Native American, Japanese, Indian, African, Heavy metal, etc. Lastly, we added ambient noise and background music to the clean speech. This resulted in $\sim$205hrs of noisy mixture for which we possess it's clean variant along with the transcript.     

\section{Experiments}
First, we tried to investigate if training with noisy and it's clean variant really helps. We trained DeepSpeech model \cite{deepspeech} as the recognizer on 100hrs of clean, noisy mixture and clean+noisy mixture. We tested the model on 5hrs of clean and noisy test dataset. 
%There are different evaluation metrics for speech enhancement techniques like SDR, SAR, etc. 
%Since we are dealing with speech enhancement in the context of ASR systems,
For evaluation, we have used the de-facto standard for ASR systems which is word error rate(WER). WER is the percentage of words mis-recognized by the ASR system(lower the better). \\
To compare our results, we have considered 3 different cases i.e., real world scenario(noisy), ideal case(noisy+clean) and our solution(noisy+enhanced). In the real world scenario, we can gather noisy data from public sources. So we trained ASR system just on noisy data. For the best case scenario, we trained DeepSpeech with noisy dataset and it's clean version. Lastly, we implemented our approach. We first processed the noisy dataset with pretrained SEGAN to get enhanced dataset. Thereafter, we trained the DeepSpeech model with noisy dataset and it's enhanced version. If our Speech enhancement model works really well, only then we might achieve similar performance as the ideal case. First two cases represent back-end approach because there is no preprocessing involved. The model is left to decide what is noise and what is not. Our approach is a hybrid of front-end and back-end approach because we clean the speech first and feed noisy and it's enhanced version for training.

\begin{figure}[ht]
\centering
\includegraphics[scale=0.35]{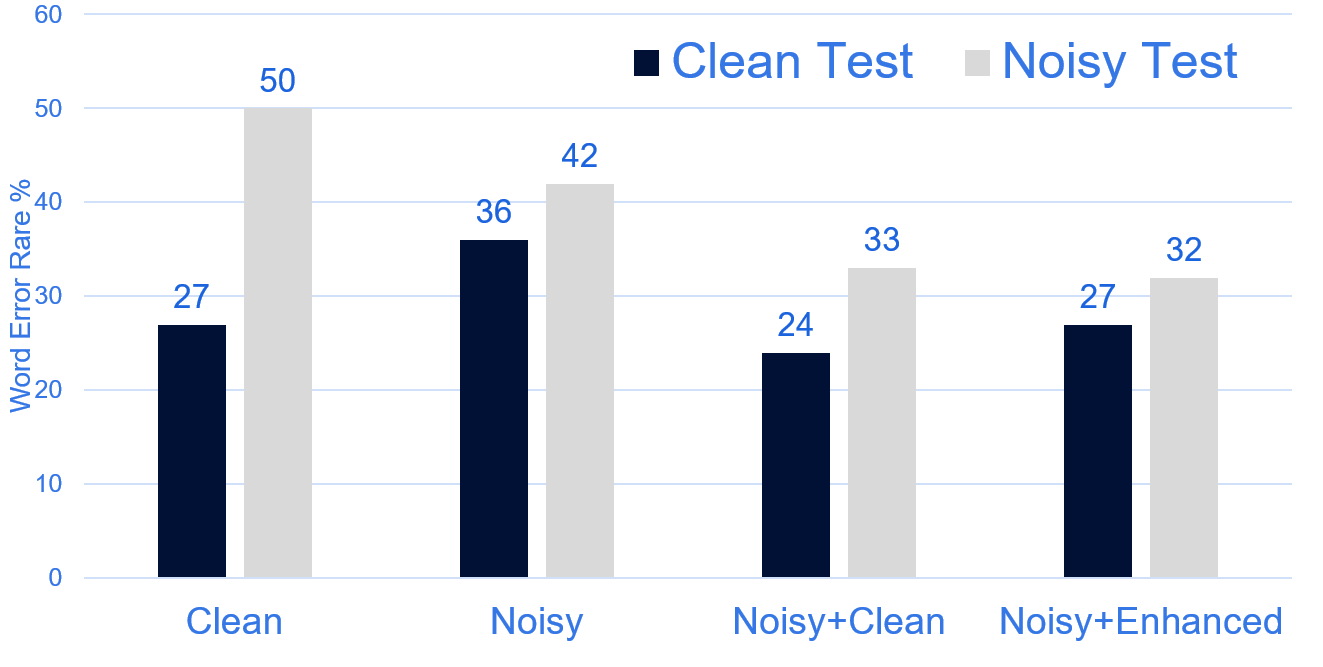}
\setlength{\abovecaptionskip}{-3pt}
\caption{Word error rate of DeepSpeech model when trained and tested on different datasets. X-axis represents different kinds of training data i.e. clean, noisy, etc. Dark colored bars represent word error rate when tested on clean data while light colored bars represents noisy test data}
\label{fig:results}
\end{figure}

\section{Results}
As shown in Fig. \ref{fig:results}, DeepSpeech model trained on clean data performs well on clean test set but lags on noisy test set. Similarly, when trained on noisy data, DeepSpeech model performs better on noisy test set but lags on clean test set. Finally, DeepSpeech model trained on clean+noisy mixture outperforms other two cases on both clean and noisy test set. %For publicly available data such as youtube, we don't have it's clean version available. 
So clearly, training with noisy and clean version helps. 

%In a multi-style ASR training, noisy speech together with its cleaned version can significantly reduce word error rate. 
Since we mightn't always have the clean version of publicly available data, we replaced clean speech with enhanced speech. Noisy speech combined with it's enhanced speech by SEGAN performed significantly well for ASR systems. We observed 9.5\% mean reduction in WER on noisy(42\% vs 32\%) and clean(36\% vs 27\%) test data when compared with real world scenario i.e. training on just noisy speech. We observed that speech cleaned with SEGAN performed at par with the ideal case scenario for noisy test dataset(32\% vs 33\%). On the clean test set, it's error rate is a little higher than ideal case(27\% vs 24\%). This might be attributed to the artifacts introduced by speech enhancement model on clean speech.      
In conclusion, this work is a proof of concept that found data treated with some speech enhancement model helps ASR become more robust and accurate.   
\section{Conclusion \& Future Work}
Our work shows that publicly available data together with Speech enhancement models can be leveraged to build robust ASR systems. 
Next, we intend to test our approach with other SE models like FSEGAN, Wave-u-net, etc. It will also be interesting to test how back-end approach compares with end-to-end approach. 
%Other possible direction will be coming up with a new deep learning framework for SE which can exploit spatial and temporal correlations in data. 
Overall, we believe this work will motivate larger research on building state of the art ASR systems from public available/found data.
  
\bibliographystyle{aaai}
\bibliography{references}

\begin{thebibliography}{}

\bibitem[\protect\citeauthoryear{Hannun \bgroup et al\mbox.\egroup
  }{2014}]{deepspeech}
Hannun, A.; Case, C.; Casper, J.; Catanzaro, B.; Diamos, G.; Elsen, E.;
  Prenger, R.; Satheesh, S.; Sengupta, S.; Coates, A.; et~al.
\newblock 2014.
\newblock Deep speech: Scaling up end-to-end speech recognition.
\newblock {\em arXiv preprint arXiv:1412.5567}.

\bibitem[\protect\citeauthoryear{Panayotov \bgroup et al\mbox.\egroup
  }{2015}]{panayotov2015librispeech}
Panayotov, V.; Chen, G.; Povey, D.; and Khudanpur, S.
\newblock 2015.
\newblock Librispeech: an asr corpus based on public domain audio books.
\newblock In {\em Acoustics, Speech and Signal Processing (ICASSP), 2015 IEEE
  International Conference on},  5206--5210.
\newblock IEEE.

\bibitem[\protect\citeauthoryear{Pascual, Bonafonte, and Serra}{2017}]{segan}
Pascual, S.; Bonafonte, A.; and Serra, J.
\newblock 2017.
\newblock Segan: Speech enhancement generative adversarial network.
\newblock {\em arXiv preprint arXiv:1703.09452}.

\bibitem[\protect\citeauthoryear{Zhang \bgroup et al\mbox.\egroup
  }{2018}]{zhang2018deep}
Zhang, Z.; Geiger, J.; Pohjalainen, J.; Mousa, A. E.-D.; Jin, W.; and Schuller,
  B.
\newblock 2018.
\newblock Deep learning for environmentally robust speech recognition: An
  overview of recent developments.
\newblock {\em ACM Transactions on Intelligent Systems and Technology (TIST)}
  9(5):49.

\end{thebibliography}

\end{document}